\documentclass[10pt,a4paper]{article}
\usepackage[utf8]{inputenc}
\usepackage[T1]{fontenc}
\usepackage{lmodern}
\usepackage[margin=2.3cm]{geometry}
\usepackage{amsmath,amssymb,booktabs,microtype}
\usepackage[hidelinks]{hyperref}
\title{\textbf{SotSort}: a stable quicksort with a tunable buffer\\[3pt]
\large Design summary and an experimental comparison (versions 8-r and 8-r5)}
\author{Lucas Sotomayor}
\date{19 November 2026 --- technical report, draft (revision 5)}
\begin{document}
\maketitle

\begin{abstract}
SotSort is a stable comparison sort whose single code path runs with anything from a fixed
16\,KB buffer to a full $n$-element buffer. Its core is a stable quicksort that partitions blocks
through the buffer and gathers them with rotations. This report summarises the design, states
which parts are borrowed, and compares an implementation against blitsort, fluxsort, glidesort,
driftsort, GrailSort and WikiSort. On one 2-vCPU machine, with 32- and 64-bit integers, the constant-memory mode takes
between 0.32 and 0.82 of blitsort's time on 17 of 22 input classes and between 0.92 and
1.07 on the other five. With a comparator that rules out the vector partition it is
slower than blitsort on 4 of 11 classes, by up to 2.3 times on sorted input. Against
Virtual-Memory Powersort, on its authors' own harness and with the same auxiliary memory, the
first measurement lost on most inputs; after three changes it located (one of them adopts that
algorithm's paging scheme), SotSort is faster in 43 of 48 type and input combinations at
$n=10^6$ on two machines. At $n=10^7$ on partially sorted integers it is between 7\% faster and
7\% slower depending on the machine, and on near-random input it is still slower: by 9\% on
integers at $n=10^7$, and by 5 to 17\% at $n=10^6$ when comparisons are expensive. All numbers are preliminary.
\end{abstract}

\section{Scope and status}
This report describes and measures one artifact, \texttt{include/sotsort.hpp}, at version
8-r5; the changes since version 8-r came from the comparison with Virtual-Memory Powersort
(Section~\ref{sec:vmps}). Two measurements in Section~\ref{sec:exp} (routing heuristics and
adversaries) are from version 8-r and were not repeated. Both commits are given in
Section~\ref{sec:avail}. The proofs (Hoare-style, in Spanish) cover
the algorithms as specified; they are not machine-checked, and the correspondence between the
proofs and this C++ file rests on differential testing (Section~\ref{sec:tests}). I have not
surveyed the literature in depth, so some ideas presented as mine may have prior art.

\section{Design}
\paragraph{Entry.} A linear scan returns on sorted input and reverses strictly descending input.
A second scan counts direction changes; if there are fewer than $n/8$, the input goes to a
natural merge sort, otherwise to the stable quicksort.

\paragraph{Stable partition.} A block of at most $B$ elements is partitioned by writing every
element both to the front of the block and to the buffer and advancing only one index (no
branch); the buffer part is then copied back. For integers with the natural order, a vector
version compares $w$ lanes against the pivot and compresses them (AVX-512 compress, or a
16-row permutation table with AVX2). Partitioned blocks $T_1F_1$ and $T_2F_2$ are fused by
rotating $F_1T_2$, in binary-counter order, so the cut point of each segment travels on a stack
and no binary search is needed.

\paragraph{Quicksort.} The pivot is a copy (a ninther up to $2^{13}$ elements, the median of
about $2n^{1/3}$ samples above), taken at pseudorandom offsets. A lower bound inherited from the
parent detects ``no element is smaller than the pivot''; the elements equal to the pivot are then
separated and are final. Unbalanced steps consume a budget of $\lfloor\lg n\rfloor+1$; when it
runs out, the range is finished by a bottom-up SymMerge sort.

\paragraph{Leaves and merges.} Ranges of at most 128 elements are split into power-of-two chunks
sorted by parity merges between the array and the buffer. A parity merge of unequal lengths $a$
and $b$ needs no bounds checks in its first $\min(a,b)$ iterations. Natural runs are merged with
the power criterion of powersort; each merge is a SymMerge whose leaves use the buffer.

\paragraph{Memory plan.} The fast mode chooses among three tiers from a byte budget: the
built-in buffer, a power-of-two block buffer, or a full buffer of $n$ elements. The budget is
the smaller of a user cap and a fraction of the available memory (including the cgroup margin).
Allocation failure steps down one tier. The same procedures run in every tier with a different
$B$; the proofs only need $B\ge 16$.

\paragraph{Routing heuristics (new in 8-r).} Three cheap probes choose among procedures
that are already proved correct, so they affect time only. (i) Before a buffered merge
on the natural-run path, eight binary searches bound the number of side changes of the
merge; long streaks get a galloping merge, otherwise a parity merge if both runs fit in
the buffer. (ii) When there are few direction changes, 256 samples at pseudorandom
positions tell whether runs have interleaved key ranges; if so, the input goes to the
quicksort unless the number of runs is small. (iii) Eight or more consecutive natural
runs shorter than eight elements are treated as a disordered zone and sorted by the
quicksort as a single run. (ii) and (iii) apply to trivially copyable types only; with
strings they measured worse.

\paragraph{What is borrowed.} The fused triple-reversal rotation (blitsort), the equal-length
parity merge (quadsort), the power criterion (powersort), galloping (Timsort), the lower-bound
treatment of duplicates (pdqsort), the budget fallback (introsort), SymMerge (Kim and Kutzner),
merge path for the parallel mode, and counting sort. \textbf{What I believe is mine}, subject to
the caveat above: the unequal-length parity merge without bounds checks, the stable vector block
partition, the unified constant-to-linear memory design with proofs parametrised by $B$, and the
three routing probes above.

\paragraph{Closest related work.} Moltmann, Nakajima and Wild~\cite{vmpowersort} attack the
same trade-off from the mergesort side: Virtual-Memory Powersort lowers the buffer of an adaptive
stable mergesort from $n/2$ to $O(\sqrt{n\log n})$ elements with the same moves and comparisons
up to an additive $O(n)$. SotSort starts instead from a constant buffer and a stable quicksort.
Section~\ref{sec:vmps} compares the two on their harness and describes the parts of their
scheme that version 8-r5 adopts. The natural-run path uses the power criterion of
Powersort~\cite{powersort}. Its galloping follows Timsort; Ghasemi, Jugé and
Khalighinejad~\cite{galloping} analyse galloping in this family of natural merge sorts. The
branch-free block partition pursues the goal of BlockQuicksort~\cite{blockqs}, avoiding branch
mispredictions, but keeps stability by writing every element to both destinations.

\section{Testing}\label{sec:tests}
Differential tests against \texttt{std::stable\_sort}, with original indices as tags to check
stability, over 15 input families: 32- and 64-bit signed and unsigned integers, pairs, records of 16, 40 and 128 bytes,
and strings; all three memory tiers; 1 to 4 threads. No failures in 19\,506 cases with AVX-512 and in 7\,410 cases per build with AVX2, no SIMD
and clang, on the final file. AddressSanitizer, UBSan and ThreadSanitizer report nothing.
Allocation failures were injected at every allocation point. Known gap: if the comparator
throws, elements can be lost.

\section{Experiments}\label{sec:exp}
\paragraph{Setup.} One cloud VM with 2 vCPUs (Intel Xeon with AVX-512), 8\,GB, Linux, gcc 13
with \texttt{-O3 -march=native}; Rust 1.97 for glidesort 0.1.2 and the standard library sorts
(\texttt{sort}: driftsort; \texttt{sort\_unstable}: ipnsort). $n=10^6$, median of 5 runs after a
warm-up, pinned to one core. Every output was checked for order, multiset and stability.
Identical runs differed by up to $\pm15\%$ on this machine. The tables and the timings in
the first paragraph below are for version 8-r5; the paragraphs on routing heuristics and
adversaries were measured with version 8-r and not repeated. The inputs: uniform random; 16 and
$\sqrt n$ distinct values; Zipf ($s=1.1$); sorted; reversed; sorted with 1\% random
replacements; sorted with a 10\% random tail; organ pipe; sorted runs of length $\sqrt n$ and
of length 63 with random keys.

\begin{table}[t]\centering\small
\begin{tabular}{lrrrr}
\toprule
 & \multicolumn{3}{c}{SotSort / blitsort} & SotSort / best of\\
Input & u32 & u64 & key+index & Grail, Wiki (32\,B)\\
\midrule
random & 0.82 & 0.71 & 1.04 & 0.56\\
16 values & 0.47 & 0.75 & 0.87 & 0.38\\
$\sqrt n$ values & 0.32 & 0.57 & 0.59 & 0.50\\
Zipf & 0.45 & 0.66 & 0.87 & 0.48\\
sorted & 0.99 & 1.07 & 2.31 & 0.52\\
reversed & 0.92 & 1.01 & 1.62 & 0.12\\
1\% noise & 1.04 & 0.63 & 0.84 & 0.50\\
10\% random tail & 0.65 & 0.72 & 0.85 & 0.75\\
organ pipe & 0.55 & 0.65 & 0.78 & 0.36\\
runs of $\sqrt n$ & 0.58 & 0.54 & 1.13 & 0.90\\
runs of 63 & 0.36 & 0.58 & 0.70 & 0.65\\
\bottomrule
\end{tabular}
\caption{Constant-memory sorts, $n=10^6$: time ratio, below 1 means SotSort is faster.
``key+index'' is a 64-bit word compared on its upper half (no vector partition). blitsort does
not support 32-byte records.}\label{tab:o1}
\end{table}

\begin{table}[t]\centering\small
\begin{tabular}{lrrrrrr}
\toprule
Input (u64) & SotSort O(1) & SotSort fast & blitsort & fluxsort & glidesort & driftsort\\
\midrule
random & 18.5 & 16.0 & 26.2 & 22.5 & 26.5 & 23.6\\
Zipf & 11.1 & 8.3 & 16.9 & 10.8 & 13.8 & 15.6\\
sorted & 0.3 & 0.4 & 0.3 & 0.3 & 0.6 & 0.5\\
reversed & 0.6 & 0.7 & 0.6 & 0.6 & 0.9 & 0.8\\
1\% noise & 8.7 & 9.9 & 13.8 & 18.8 & 26.3 & 16.8\\
10\% random tail & 4.8 & 6.5 & 6.6 & 6.4 & 4.4 & 5.8\\
organ pipe & 2.4 & 3.3 & 3.8 & 2.6 & 1.6 & 4.0\\
runs of $\sqrt n$ & 16.9 & 23.6 & 31.2 & 21.5 & 12.2 & 39.9\\
runs of 63 & 24.2 & 17.4 & 41.4 & 23.8 & 26.9 & 20.3\\
\bottomrule
\end{tabular}
\caption{Milliseconds, $n=10^6$ 64-bit integers, one thread. ``fast'' uses a full buffer. The
rows with 16 and $\sqrt n$ values are omitted here because the fast mode answers them with a
counting sort.}\label{tab:abs}
\end{table}

\paragraph{Reading.} With 32- and 64-bit integers the constant-memory mode is ahead of
blitsort or level with it on every input class measured; the closest are sorted,
reversed and nearly sorted input (0.92 to 1.07). With a comparator that the vector
partition cannot use (``key+index'') it loses on 4 of 11 classes, worst on sorted input
(2.31), where blitsort's scan is faster. Against the linear-memory sorts, the fast mode
is ahead on random input and behind glidesort on organ pipe, the random tail and runs of
$\sqrt n$ (1.5 to 2.1 times for 64-bit integers). For $n=10^7$ random 64-bit integers
(median of 3) the fast mode took 200\,ms, the constant-memory mode 287, glidesort 294,
fluxsort 311, driftsort 332 and blitsort 462. Extra memory at $n=10^7$ (80\,MB of data), as
peak resident set growth: under 0.5\,MB in constant-memory mode and about 37\,MB in fast mode.
Absolute times are lower than in revision 4 of this report in most cells, including for the
unchanged competitors (blitsort, random \texttt{u64}, $n=10^6$: 32.6 then 26.2\,ms), so the
shared VM was faster on this day; compare within a table, not across revisions.

\paragraph{Effect of the routing heuristics.} Against the same code without them
(interleaved runs, minimum of 7): 0.45 to 0.81 of the time on sorted data with a random
zone, 0.26 to 0.88 on sorted runs of random keys, no change on random and
duplicate-heavy input, and at most 10\% worse elsewhere, which is within noise. Buffer
sizes from 4 to 128\,KB and leaf sizes from 32 to 256 made no measurable difference.

\paragraph{Comparisons.} On random input every sort measured makes between 1.09 and 1.22 times
$\lg n!$ comparisons; SotSort makes 1.10. This is not a differentiator.

\paragraph{Adversaries.} McIlroy's adversary, which fixes values lazily, loses to any sort with
a run-detection pass, so a random fraction ($10^{-3}$ to $10^{-1}$) of positions was frozen in
advance. The largest count over $n=10^5,10^6$ was $1.14\,n\lg n$ for SotSort, $0.65$ for
blitsort, $1.96$ for driftsort and $3.67$ for ipnsort. No sort showed quadratic behaviour; this
is evidence, not a proof. A second, lazy adversary that gives every pair of undefined elements
the same minimal value was run against the quicksort alone. With the budget charge on the
equal-separation step removed (the state of version 6), it forces 167, 335, 668 and 1335
comparisons per element for $n=1000$ to $8000$, about $n/6$; with the charge, 20 to 27. That
missing charge was found while writing the termination proof, after fuzzing had not shown it.

\section{Comparison with Virtual-Memory Powersort}\label{sec:vmps}
\paragraph{Setup.} I added SotSort to the experimental harness published with
\cite{vmpowersort} (branch \texttt{assignment-fix}, which also counts copy and move
constructions) and changed nothing else: their inputs (runs of geometrically distributed
length with mean $S$, keys random within a run), their four element types (\texttt{int}, cheap
to compare and to move; \texttt{ptr}, a pointer to 30 integers of which only the last is
nonzero, so comparisons are slow and moves cheap; \texttt{arr}, the same 30 integers held by
value, slow in both; \texttt{fastarr}, 30 random integers by value, cheap comparison and slow
moves), their timing with \texttt{clock()}, their correctness check and \texttt{-O2}. Their
wrapper type is not trivially copyable, so SotSort runs its generic path: no SIMD, no routing
probes, no disordered zones. SotSort ran single-threaded with \emph{the same memory}: the
\texttt{max\_bytes} given to SotSort is exactly what Virtual-Memory Powersort allocates, and
SotSort's plan counts against it everything it allocates (the buffer as rounded by the
allocator, its internal 16\,KB buffer and its page tables); a test checks the peak.

\paragraph{What the comparison changed.} The first measurement (version 8-r) lost on most
inputs. Counting operations located three causes, fixed in versions 8-r3 to 8-r5.
(1) The routing counted every turn before the run detection scanned the array again, one
extra comparison per element on presorted input; 8-r3 estimates the turn density from 64
windows of 256 pairs when $n\ge 2^{16}$. (2) All extra moves came from rotations in merges
whose runs both exceed the buffer: 3.75--4.30 moves per element at $n\approx 10^6$, against
0.71--0.75 with $n/2$ memory. 8-r4 replaces them by a page merge: the range is cut into pages
of $B/8$ elements, the buffer supplies 8 spare pages, output pages go to any free page and a
final cycle-following pass puts them in order. (3) Merges that fit in the buffer still cost
about 1.5 moves per element; 8-r5 adopts the scheme of \cite{vmpowersort} in the natural-run
path when the buffer is at most $n/8$ and the first 17 detected runs average at least 32
elements: runs are copied to pages on detection, merged page to page, and moved back once.
Both the page merge and the paged runs are the authors' idea; what SotSort adds around them is
its run detection, the quicksort for near-random input, and the choice between paths. All
three changes only select or reorganise stable merges, so correctness rests on the same
arguments; the test suite (about 13\,500 differential cases, sanitizers, and a memory-peak test) passes.
A last problem was code generation: with gcc \texttt{-O2} the larger functions had left the
comparator or the insertion loop out of line, which cost up to 30\% with expensive
comparisons; it was found because operation counts were identical while times were not.

\begin{table}[t]\centering\small
\begin{tabular}{@{}lrrrrrr@{}}\toprule
$S$ & 2 & 100 & 1000 & $10^4$ & $10^5$ & $10^6$\\\midrule
time, SotSort / VM-Powersort, cloud & 1.09 & 0.97 & 0.96 & 0.97 & 1.04 & 1.07\\
time, SotSort / VM-Powersort, PC & 1.09 & 0.95 & 0.93 & 0.93 & 0.95 & 0.99\\
comparisons per element, VM-Powersort & 26.80 & 17.26 & 13.58 & 10.52 & 7.24 & 3.98\\
comparisons per element, SotSort & 22.30 & 16.93 & 13.58 & 10.52 & 7.24 & 3.98\\
moves per element, VM-Powersort & 30.71 & 19.37 & 15.55 & 12.50 & 9.22 & 5.96\\
moves per element, SotSort & 49.58 & 18.69 & 15.31 & 12.27 & 8.99 & 5.96\\\bottomrule
\end{tabular}
\caption{Version 8-r5, \texttt{int}, same memory, $n$ uniform in $[0.9\cdot10^7, 10^7]$
(per-element counts divided by $0.95\cdot10^7$), 10 runs for time on each machine, 3 for counts (counts are the same on both).
``Moves'' counts assignments and copy/move constructions.}\label{tab:vm7}
\end{table}

\begin{table}[t]\centering\small
\begin{tabular}{@{}llrrrrrr@{}}\toprule
machine & type & $S=2$ & 100 & 1000 & $10^4$ & $10^5$ & $10^6$\\\midrule
cloud & \texttt{int}     & 1.01 & 0.88 & 0.88 & 0.89 & 1.00 & 0.78\\
cloud & \texttt{ptr}     & 1.05 & 0.79 & 0.82 & 0.80 & 0.93 & 0.97\\
cloud & \texttt{fastarr} & 1.10 & 0.79 & 0.89 & 0.88 & 0.90 & 0.57\\
cloud & \texttt{arr}     & 0.80 & 0.84 & 0.85 & 0.96 & 0.94 & 0.77\\
PC & \texttt{int}     & 0.99 & 0.91 & 0.89 & 0.83 & 0.98 & 0.72\\
PC & \texttt{ptr}     & 1.17 & 0.68 & 0.63 & 0.71 & 0.92 & 0.85\\
PC & \texttt{fastarr} & 0.97 & 0.84 & 0.84 & 0.87 & 0.89 & 0.52\\
PC & \texttt{arr}     & 0.67 & 0.79 & 0.84 & 0.82 & 0.66 & 0.58\\\bottomrule
\end{tabular}
\caption{Version 8-r5, time SotSort / Virtual-Memory Powersort with the same memory,
$n\approx10^6$, 10 runs. Cloud: the 2-vCPU VM of Section~\ref{sec:exp}; PC: a 2-vCPU Linux VM
on an Intel Core Ultra 7 155H, gcc 11.4. Same source on both.}\label{tab:vm6}
\end{table}

\paragraph{Results.} At $n\approx10^6$ (Table~\ref{tab:vm6}) the ratio is below 1.00 in 43 of
48 cells, down to 0.52 for slow moves with long runs. At $n=10^7$ on integers
(Table~\ref{tab:vm7}) SotSort is 1--7\% faster on the PC for $S\ge100$; on the cloud VM it is
3--4\% faster for $S$ from 100 to $10^4$ and 4--7\% slower for $S=10^5$ and $10^6$, where its
operation counts are equal or lower. Since that loss does not appear on the second machine I
read it as a property of that VM rather than of the algorithm, without being able to show why
(no hardware counters). On both machines SotSort is 9\% slower at $S=2$. The weak spot is near-random input ($S=2$): SotSort's quicksort makes
17--32\% fewer comparisons but many more moves (49.58 against 30.71 per element at $n=10^7$,
since with this budget its partition blocks are small), so it loses when moves or comparisons
are not both cheap: \texttt{ptr} 1.05 and 1.17, cloud \texttt{fastarr} 1.10. With $n/2$
memory against CPython-style Powersort SotSort is still 1--11\% slower on \texttt{int} at
$n=10^7$ on the cloud VM (6--9\% on the PC) and up to 1.56 times slower on \texttt{ptr} at $n=10^6$; that comparison was not the
target of these changes. Repeated runs of identical code moved ratios by up to about 10\%, so
smaller differences are ties.

\paragraph{An observation about the baselines.} My absolute times for Virtual-Memory
Powersort are close to those the authors report for an i7-4770 in the data published with the
code (\texttt{int}, $n=10^7$: 0.128\,s there and 0.128 to 0.129\,s on the cloud VM here for
$S=10^6$; 0.82 and 0.93\,s for $S=2$). In those data copy-Powersort is about 10--25\% faster
than Virtual-Memory Powersort; on my two machines, at $n=10^7$,
the two are within about 5\%. I have not found the cause. Candidates are the processor
(their i7-4770 against a recent Xeon and a Core Ultra 7, both behind a hypervisor), the
compiler version, and differences between their published runs and the \texttt{assignment-fix}
branch I used. This matters for reading Table~\ref{tab:vm6}: the comparison here is against
Virtual-Memory Powersort as it runs on these machines.

\section{Limitations}
For Section~\ref{sec:exp}, one machine, virtualised and shared, with two cores (Section~\ref{sec:vmps}
adds a second virtual machine, still no bare metal); no hardware counters; no ARM; no strings in
the comparison tables; the competitors were built with their default settings; glidesort and
driftsort cross a C-to-Rust call boundary once per sort. The comparison against blitsort uses
its default 512-element buffer, while SotSort's buffer is 16\,KB (2048 64-bit elements): both
are constant, but they are not equal. The thresholds
of the routing probes were tuned on this one machine, on some of the same inputs used for the
comparison, so the structured-input rows are optimistic.

\section{Availability}\label{sec:avail}
Code, tests, the benchmark harness and raw measurements: \url{https://github.com/sotomayorlucas/sotsort},
code at commit \texttt{c16d43f} for version 8-r and \texttt{b9e0df1} for version 8-r5 (MPL-2.0).
The Virtual-Memory Powersort comparison is in \texttt{bench/vmpowersort}: a patch to their
harness, the scripts and the raw results; their code is not redistributed.

\end{document}